\documentclass[10pt,conference]{IEEEtran}
\usepackage[utf8]{inputenc}

\usepackage{booktabs}
\usepackage{hyperref}
\usepackage{amsmath}
\usepackage{enumerate}
\usepackage[shortlabels]{enumitem}
\usepackage{url}
\usepackage{graphicx}
\usepackage{setspace}
\usepackage{xspace}
\usepackage{multirow}
\usepackage{balance}
\usepackage[noadjust]{cite}

\newcommand{\dataset}{\textit{SlowOps}\xspace}

\begin{document}

\title{Stack Trace Deduplication: Faster, More Accurately, and in More Realistic Scenarios}

\author{\IEEEauthorblockN{Egor Shibaev\IEEEauthorrefmark{4}\IEEEauthorrefmark{2}, Denis Sushentsev\IEEEauthorrefmark{1}, Yaroslav Golubev\IEEEauthorrefmark{2}, Aleksandr Khvorov\IEEEauthorrefmark{4}\IEEEauthorrefmark{1}}
\IEEEauthorblockA{\IEEEauthorrefmark{1}\textit{JetBrains}, \IEEEauthorrefmark{2}\textit{JetBrains Research}, \IEEEauthorrefmark{4}\textit{Constructor University} \\
eshibaev@constructor.university, \{denis.sushentsev, yaroslav.golubev, aleksandr.khvorov\}@jetbrains.com
}
}

\maketitle

\begin{abstract}

In large-scale software systems, there are often no fully-fledged bug reports with human-written descriptions when an error occurs. In this case, developers rely on stack traces, \textit{i.e.}, series of function calls that led to the error. Since there can be tens and hundreds of thousands of them describing the same issue from different users, automatic deduplication into categories is necessary to allow for processing. Recent works have proposed powerful deep learning-based approaches for this, but they are evaluated and compared in isolation from real-life workflows, and it is not clear whether they will actually work well at scale.

To overcome this gap, this work presents three main contributions: a novel model, an industry-based dataset, and a multi-faceted evaluation. Our \textit{model} consists of two parts --- (1) an embedding model with byte-pair encoding and approximate nearest neighbor search to quickly find the most relevant stack traces to the incoming one, and (2) a reranker that re-ranks the most fitting stack traces, taking into account the repeated frames between them. To complement the existing datasets collected from open-source projects, we share with the community \dataset~--- a \textit{dataset} of stack traces from IntelliJ-based products developed by JetBrains, which has an order of magnitude more stack traces per category. Finally, we carry out an \textit{evaluation} that strives to be realistic: measuring not only the accuracy of categorization, but also the operation time and the ability to create new categories. The evaluation shows that our model strikes a good balance --- it outperforms other models on both open-source datasets and \dataset, while also being faster on time than most. We release all of our code and data, and hope that our work can pave the way to further practice-oriented research in the area.

\end{abstract}


\section{Introduction}

In major software companies, developers receive a lot of error reports from individual users. While detailed bug reports can help developers fix the issues, users often do not have the time or the desire to write them~\cite{bugreps2011}. In this case, software projects often rely on \textit{automatic bug reports}~\cite{automaticbugreps2010}, which in turn rely heavily on \textit{stack traces}, \textit{i.e.}, stacks of method calls called during the error, accompanied by various metadata~\cite{stacktraceforreports2004}.

Different bug reports often relate to the same error, and since there are tens and hundreds of thousands of them in large companies, an automatic \textit{deduplication}~\cite{duplikates2008} is necessary, which involves putting the new incoming report into one of the existing categories with similar reports. In the case when bug reports are represented by stack traces, this requires a \textit{similarity measure} to compare them~\cite{automaticdeduplication2007}.

This problem is well-studied in literature, and there exist numerous solutions~\cite{editdis, editdis6, editdis7, 6100061, DURFEX, lerch, tracesim, khvorov2021s3m, karasov2022aggregation, deepcrach, fast}. The simplest methods are string-based, using different techniques such as Levenshtein distance, prefix match, etc.~\cite{editdis, editdis6, editdis7, tracesim}. Also popular are information retrieval-based approaches~\cite{lerch, tracesim, DURFEX}, with one of the most popular ones developed by Lerch and Mezini~\cite{lerch}, employing the TF-IDF measure to find similar reports. With the rise of deep learning, new solutions were introduced that utilized it. Khvorov et al. proposed the first such model called S3M~\cite{khvorov2021s3m}, which used a biLSTM~\cite{lstm} architecture with aggregation on top of outputs to encode stack traces and a linear transformation on top to compare two stack traces. Subsequently, Liu et al.~\cite{deepcrach} improved upon this approach to achieve better results.
 
Despite these advancements, there are still significant challenges that need to be addressed. Firstly, the current methods still do not provide sufficient accuracy for a completely automatic usage in real projects, requiring further enhancement. The accuracy is critical to ensure that developers can rely on these systems. Current state-of-the-art deep learning methods often process each stack frame individually, and only afterwards are the embeddings compared. To address this issue, the use of a reranker --- a model that processes two entities simultaneously --- may be reasonable, as it is often used in retrieval tasks to enhance the precision of the results~\cite{Nogueira2019PassageRW}.

Also, better evaluation methods are critical. Firstly, the evaluations are limited by the narrow scope of existing open-source datasets~\cite{campbell2016unreasonable, rodrigues2022tracesim}, which affects their generalizability and robustness. Secondly, the vast majority of existing works only measure the general accuracy of putting reports into the necessary category, without considering two aspects crucial for practical use in industry: the ability to correctly create new categories and the operation time. To the best of our knowledge, only two works~\cite{deepcrach, fast} studied these aspects. However, their analysis of time did not account for the possibility to pre-compute and cache embeddings, which is commonly done with embedding-based systems in large software projects. As for the evaluation of creating new categories, it was carried out before the advent of modern deep learning-based approaches, and it is thus necessary to also compare them on this task.

In this paper, we aim to fill all these gaps in research by developing a new model that performs better than the current state of the art, presenting a novel dataset collected from large-scale industrial products, and carrying out a more comprehensive, multi-faceted evaluation. Our model consists of two stages. The first stage is a biLSTM-based embedding model designed to quickly encode stack traces into compact feature vectors. The main features of this stage are Byte Pair Encoding (BPE)~\cite{sennrich-etal-2016-neural} and the faster approximate nearest neighbours search~\cite{faiss} to find top-\textit{k} closest stack traces to the incoming one. The second stage is a reranker, which is used to more accurately re-rank the selected top-\textit{k} stack traces further. This reranker is also based on a biLSTM architecture but differs in that it processes two stack traces simultaneously, specifically focusing on stack frames that are repeated across them. This method allows for a more nuanced comparison than is impossible with embeddings alone, as it directly addresses the interactions between the elements of the stack traces, thus increasing the accuracy of the similarity score.

To test the new model, as well as existing solutions, in a more exhaustive and real-world manner, we conducted a new evaluation. Firstly, we collected \dataset~--- a novel industrial dataset from JetBrains, a large vendor of software for developers and teams. This dataset includes error reports caused by \textit{Slow Operation Assertion} in IntelliJ-based products~\cite{kurbatova2021intellij} and features an order of magnitude more reports per category compared to the existing open-source datasets, providing a distinct source of data for evaluation.

Using \dataset, as well as existing datasets, we compared the models not only in terms of general accuracy, but also measured the correctness of creating new categories and the time taken to determine the suitable category for a new error report, emphasizing the importance of speed. For the time comparison, we only compared the time necessary to compute the similarity scores, assuming the pre-computed embeddings of existing stack traces, as is commonly done in large systems (including at JetBrains). For our two-stage solution, we measured both the time of retrieval using the embedding model and the reranking time.  

The results show that our model outperforms existing methods across multiple dimensions. In terms of the accuracy of classifying stack traces, our approach outperforms other models on all datasets. For creating new categories, our approach also outperforms other models on all datasets except for \dataset, where it ties with an LLM text-embedding-3-small~\cite{openai2022textembedding}. In both tasks, the next best model is the LLM, also showing very good results. The results on \dataset differ greatly from open-source datasets, highlighting the importance of carrying out evaluations on diverse data. At the same time, models show very different results in terms of speed of operation. While the well-performing LLM takes 1021.2 ms on average, our model takes 144.5 ms with reranker and just 8.7 ms without it. It can be seen that the proposed approach represents a balanced combination of accuracy and efficiency. 

Our dataset, \dataset, is publicly available on Zenodo~\cite{dataset}. The code for our model and the conducted evaluation is available in the replication package on GitHub~\cite{replication}.

Overall, the main contributions of this paper are as follows:

\begin{itemize}
    \item \textbf{State-of-the art solution}. We proposed a novel approach for deduplicating stack traces that includes an embedding model and a reranker. This architecture achieves an Acc@1 score of 0.52 on the most popular public NetBeans dataset, which is higher than previous solutions. 
    
    \item \textbf{Dataset}. We introduced \dataset, a dataset based on the private data of JetBrains, a large software company. The dataset has an order of magnitude more reports per category than existing open-source datasets and demonstrates different results of evaluation, making it a valuable addition from the industry.

    \item \textbf{Realistic evaluation}. We conducted a multi-faceted evaluation: the accuracy of categorization, the ROC-AUC score to evaluate the model's ability to create new categories, and time to find the most suitable category. Our model shows improvements in both speed and the creation of new categories, making it highly effective for real-world applications.

\end{itemize}

The rest of the paper is organized as follows. In Section~\ref{sec:background}, we introduce the necessary concepts and provide an overview of the existing work in the field. In Section~\ref{sec:methodology}, we provide the description of our new approach. Section~\ref{sec:evaluation} describes the novel dataset we collected, the evaluation settings, and the results of our experiments. In Section~\ref{sec:discussion}, we discuss these results. Finally, Section~\ref{sec:ttv} provides the threats to the validity of our work and Section~\ref{sec:conclusion} concludes it.
\section{Background \& Related Work}
\label{sec:background}

A \textbf{stack trace} is a crucial component of an error report in software development, depicting the sequence of function calls that led to an error or an exception. It consists of a series of \textbf{stack frames}, each representing a specific function call within the source code. A stack trace provides a systematic breakdown of the operations leading to a failure, allowing developers to trace back through the execution process.

Even though stack trace is only a part of an error report, most of the described approaches, as well as our one, do not utilize other metadata. Because of this, in this paper, we sometimes use the terms \textit{stack trace}, \textit{bug report}, \textit{error report}, and just \textit{report} interchangeably for easier presentation. 

\subsection{Similarity Models for Stack Trace Deduplication}

The problem of searching for similar reports based on stack traces has been approached with non-deep learning methods since the 2000s. Most of these approaches use string matching algorithms. A notable method outlined by Brodie et al. \cite{biologseq} adopts a biological sequence searching algorithm, specifically, a modified Needleman-Wunsch algorithm~\cite{needleman1970general}. This method involves preprocessing the stack trace by removing typical error-handling and common entry routines, as well as eliminating recursive function calls.

Modani et al. \cite{editdis} explore various methods for comparing stack traces, including techniques based on edit distance, the longest common subsequence, and prefix matching. Additionally, they introduce an indexing strategy for all available stack traces to expedite the search process.

Related techniques are discussed by Bartz et al. \cite{editdis6} and by Dhaliwal et al. \cite{editdis7}, with the latter being particularly noteworthy. This method employs a two-step process where signatures are first created for each stack trace, followed by the computation of Levenshtein distance between these signatures to determine similarity. This combination enhances the precision and efficiency of stack trace comparison.

Lerch and Mezini \cite{lerch} depart from traditional string matching algorithms, utilizing information retrieval techniques instead. Specifically, they employ TF-IDF scores to evaluate the similarity between two stack states. The similarity score between a query $q$ and a document $d$ is computed as follows (here, both the document and the query represent stack traces):

$$
\text{score}(q, d) = \sum_{f \in q} \text{tf}_d(f) \cdot \text{idf}(f)^2
$$

\noindent where $tf$ is term frequency and $idf$ is inverse document frequency of frame $f$.

The DURFEX method, introduced by Sabor et al. \cite{DURFEX}, treats stack traces as sequences of package names. This technique segments each sequence into multiple N-grams and maps them into fixed-size sparse feature vectors. The advantage of DURFEX is its scalability, effectively handling large volumes of bug reports typically encountered in production environments. This approach helps mitigate challenges related to the management and analysis of extensive error report data.

Both string matching algorithms and information retrieval techniques have shown success independently. Therefore, a hybrid approach that combines these paradigms was considered promising. The method described by Vasiliev et al. \cite{tracesim} integrates Levenshtein distance for direct string comparison with the weighting of frames based on their IDF scores. By merging these two successful techniques, this method has achieved impressive results.

The recent paper by Rodrigues et al. \cite{fast} introduces the FaST method, which addresses the high time complexity often required to compare stack traces. Unlike traditional techniques that operate in quadratic time, FaST functions in linear time, making it better suited for handling large volumes of reports. This efficiency is achieved through a heuristic that aligns frames closest to the top of the stack, based on the observation that these upper frames are typically more relevant.

The first deep learning approach for stack trace comparison, called S3M, was introduced by Khvorov et al. \cite{khvorov2021s3m}. This method begins by trimming and tokenizing each frame of a stack trace, which reduces variability. Each tokenized frame is then processed using a biLSTM encoder to generate hidden representations, which are concatenated to form embeddings for each stack trace. A multilayer perceptron (MLP) then calculates a similarity score based on Euclidean distance, sum, and pointwise product of embeddings. The model was trained on triplets (anchor, positive example, negative example) using RankNet loss to enhance ranking accuracy, significantly outperforming previous non-deep learning methods. However, the method has limitations, such as the potential undertraining of embeddings for rare functions and inefficiency, as it requires running a neural network for each stack trace pair rather than using an indexing system like FAISS \cite{faiss} for quicker retrieval.

The DeepCrash method~\cite{deepcrach} utilizes deep learning to enhance stack frame encoding through its Frame2vec framework. This model splits each frame into subframes or package names, and employs a skip-gram model \cite{skipgram} to train embeddings for these subframes. The frame’s embedding is then derived from the mean of its subframes’ embeddings, and a biLSTM is used to generate the complete stack trace embedding. Although DeepCrash effectively reduces vocabulary size by segmenting frames into subframes, it struggles with new packages added to the repository because embeddings for corresponding tokens would not be initialized.

Overall, it can be seen that deep learning-based approaches represent the current state of the art, however, they can and should still be improved for the actual use in practice.

\subsection{Evaluation of Similarity Models}

\begin{figure*}[h]
    \centering
    \includegraphics[width=\textwidth]{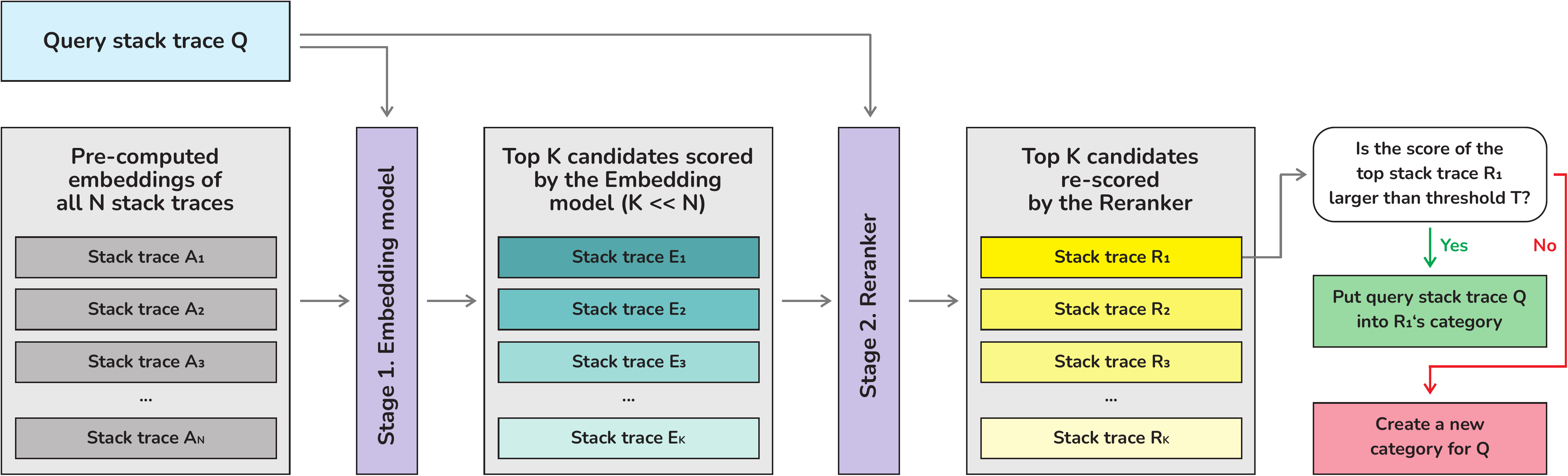}
    \caption{The general pipeline of the proposed approach.}
    \vspace{-0.3cm}
    \label{fig:rerank}
\end{figure*}

The majority of evaluations presented in previous works use the following public datasets: Ubuntu, Eclipse, NetBeans, and Gnome, --- published by Campbell et al.~\cite{campbell2016unreasonable} and Rodrigues et al.~\cite{rodrigues2022tracesim}. Since we use all of them for evaluation, a more detailed description of these datasets is provided further in Section \ref{subsec:exdatasets}. Works by Chao et al. \cite{deepcrach} and Khvorov et al. \cite{khvorov2021s3m} used private datasets that are not publicly available.

Previous methods measured the accuracy of categorization in two main ways. The first method was to measure retrieval metrics such as Mean Reciprocal Rank (MRR) or Recall Rate at k (RR@k). In this case, all events that involve attaching a report to a category are sorted by their arrival time. A similarity model is then used to retrieve the most relevant categories for each event, and metrics are computed based on the placement of the true category. Another method of measuring model accuracy is to compute clustering metrics such as Purity and Inverse Purity. These metrics assess how well a model can create categories without any labels. One of the works that utilize this approach is by Chao et al. \cite{deepcrach}.

We argue that simply measuring the accuracy of categorization is not enough to evaluate the ability of a model to work in practice. A crucial work in this regard is the work of Rodrigues et al.~\cite{fast} --- it evaluated the correctness of creating new categories by employing the ROC-AUC metric and also evaluated the time complexity. However, its comparison of the creation of new categories was carried before the release of state-of-the-art deep learning models, and so it is of interest to repeat it for them. As for the time complexity, this addresses the problem but does not study it exhaustively, because the authors did not actually measure the time. The time was directly measured in the work of Chao et al.~\cite{deepcrach}, however, in a different setting. There, the authors evaluate clustering approaches, and therefore measure the time with computing the embeddings of all the stack traces. Considering the fact that embeddings are often pre-computed, this might not indicate the actual performance of models in realistic settings.

Our paper addresses these issues by providing a multi-faceted evaluation. We provide a new dataset based on closed data from JetBrains and, in addition to general accuracy, evaluate the ability to create new categories and operation time. In all our measurements, we consider the latest state-of-the-art deep learning models, and also compare specifically the time to calculate the similarity metric to account for pre-computed embeddings used in deep learning approaches.
\section{Approach}
\label{sec:methodology}

In real-world scenarios, when a new error report arrives, the system for report grouping needs to find the most relevant category for the given report or create a new category. This is a retrieval task. To address this, we propose a two-stage approach: retrieval using an embedding model followed by more accurate re-ranking using a more advanced model, a well-known approach in retrieval tasks. The overall pipeline of the proposed approach is presented in Figure~\ref{fig:rerank}.

The advantage of the embedding model is that embeddings of the already arrived reports can be precomputed and stored in an index such as FAISS~\cite{faiss}. Then, when a new error report arrives, only its embedding needs to be computed. After the most relevant candidates are quickly retrieved, the reranker can be used to rank them more accurately. This way, the system combines the speed of the embedding model with the accuracy of the reranker. Let us now describe each of the two stages in greater detail.

\subsection{Embedding Model}
\label{sec:embedding_model}

The first part of our approach is the embedding model, the goal of which is to process all \textit{N} pre-computed embeddings of the existing stack traces and select \textit{K} most perspective ones that are similar to the incoming one, so that $K \ll N$.

\subsubsection{Preprocessing and tokenization of stack frames}

In the S3M model~\cite{khvorov2021s3m}, each frame is assigned an embedding, and in the DeepCrash model~\cite{deepcrach}, each package is assigned an embedding. This can cause problems as new frames and packages may be added to the repository and their embeddings will be initialized randomly. To handle this problem, we use Byte Pair Encoding (BPE)~\cite{stacktraceforreports2004} to split package names into tokens. With BPE, if a new package is added, its name will be split into tokens known to the model, and it also allows for a pre-determined vocabulary size.  The preprocessing workflow is streamlined as follows:

\begin{enumerate}
  \item Split each string into package, class, and method names.
  \item Further split each package name using camelCase splitting, suitable for Java and Kotlin projects. This method can be adjusted for other languages, \textit{e.g.}, in Python, one might split based on underscores (snake\_case).
  \item Apply BPE to each resulting string, converting sequences into tokens and mapping each to a token index.
\end{enumerate}

The BPE tokenizer is trained using the same dataset as the model. This tokenization procedure enables controlled vocabulary sizing, setting the number of tokens prior to training. In this study, the limit was set to 10,000 tokens, as selected in preliminary experiments, and we leave further experiments for future work. Training BPE on the same dataset makes this tokenization more dataset-specific, enhancing the relevance of the generated tokens for the given source of reports. This is also the reason for training a new BPE instead of fine-tuning an existing one, since stack traces contain a lot of unique terminology, and the datasets provide enough data.

\subsubsection{Creating embeddings of stack frames}

Once each stack frame is tokenized, it can be converted into a vector representation. While initial methods like DeepCrash used a bag-of-words (BOW) model, averaging embeddings of all tokens in a frame, we considered more advanced techniques for this transformation, similar to the work by Pradel et al. \cite{Pradel2020ScaffleBL}. Since further on, similar to previous works, we use a bi-directional LSTM (biLSTM) to combine the embeddings of frames into the embedding of an entire stack trace (see next Section~\ref{sec:bilstm}), we decided to try the same approach to combine the embeddings of tokens into the embedding of a frame on this previous step. It showed a slightly better result, so we decided to use it. Since it is more critical in the next step, we will describe it in detail for combining the embeddings of frames into the embedding of an entire stack trace.

\subsubsection{Creating embeddings of stack traces}
\label{sec:bilstm}

After converting each stack frame into a vector, the next step involves computing the embedding for the entire stack trace, represented as a sequence of vectors. Like we just mentioned, we used a bidirectional LSTM (biLSTM) to encode these sequences, following recent state-of-the-art approaches~\cite{khvorov2021s3m, deepcrach}.

In our biLSTM model, each input frame's embedding is processed bidirectionally, generating a pair of output vectors \(o_i\) (one for each direction) for each frame, which are then concatenated. Additionally, the final hidden states from both directions are concatenated to form \(h\), representing the overall context of the sequence. To aggregate these outputs into a single vector representation of the stack trace, we explored the following methods:

\begin{enumerate}
    \item \textbf{Average}: Compute the mean of all output vectors.
    $$agg_{avg} = \frac{1}{n} \sum_{i=1}^{n} o_i$$
    \item \textbf{Max}: Determine the maximum value across each dimension of the output vectors.
    $$agg_{max}^{(j)} = \max(o_1^j, ..., o_n^j)$$
    \item \textbf{Hidden}: Use the hidden state as the embedding.
    $$agg_{hidden} = h$$
\end{enumerate}

The most effective aggregation method combined these three strategies, concatenating the average, max, and hidden state embeddings into a single vector, yielding the best performance on our datasets.

\subsubsection{Training the Embedding model}
\label{sec:embedding_mode_training}

The training data consisted of pairs of stack traces: an anchor stack trace and a positive example (a stack trace from the same category). For each category, we sampled positive pairs. We set a parameter \textit{max\_pairs\_per\_category} for each dataset separately, depending on the average number of unique reports. If the number of possible pairs is lower than this value, then all pairs are included. Otherwise, a random subset of \textit{max\_pairs\_per\_category} pairs is sampled from the category, ensuring that no single category dominates the training dataset, which could lead to overfitting. The training was carried out in batches, with negative examples for an anchor stack trace being selected from all other pairs in the same batch.

We utilized the InfoNCE loss function to optimize our model, a widely recognized choice in unsupervised learning frameworks \cite{infoince1, infoince2, infoince3}. The InfoNCE loss is defined as:
\[
L_{\text{InfoNCE}} = -\log \frac{\exp(s_p / \tau)}{\sum_{k=1}^{N - 1} \exp(s_n^{(k)} / \tau)}
\]
where \(s_p\) is the similarity score between the anchor and the positive example, \(s_n^{(k)}\) are the similarity scores between the anchor and each of the \(N - 1\) negative examples within the batch, where \(N\) is the batch size. \( \tau \) is a temperature parameter that adjusts the distribution of the scores. In each batch, each anchor report has one positive and \(N - 1\) negative examples, preventing the model from merely memorizing stack traces.

\subsection{Reranker}
\label{sec:reranker}

\begin{figure*}[h]
    \centering
    \includegraphics[width=0.8\textwidth]{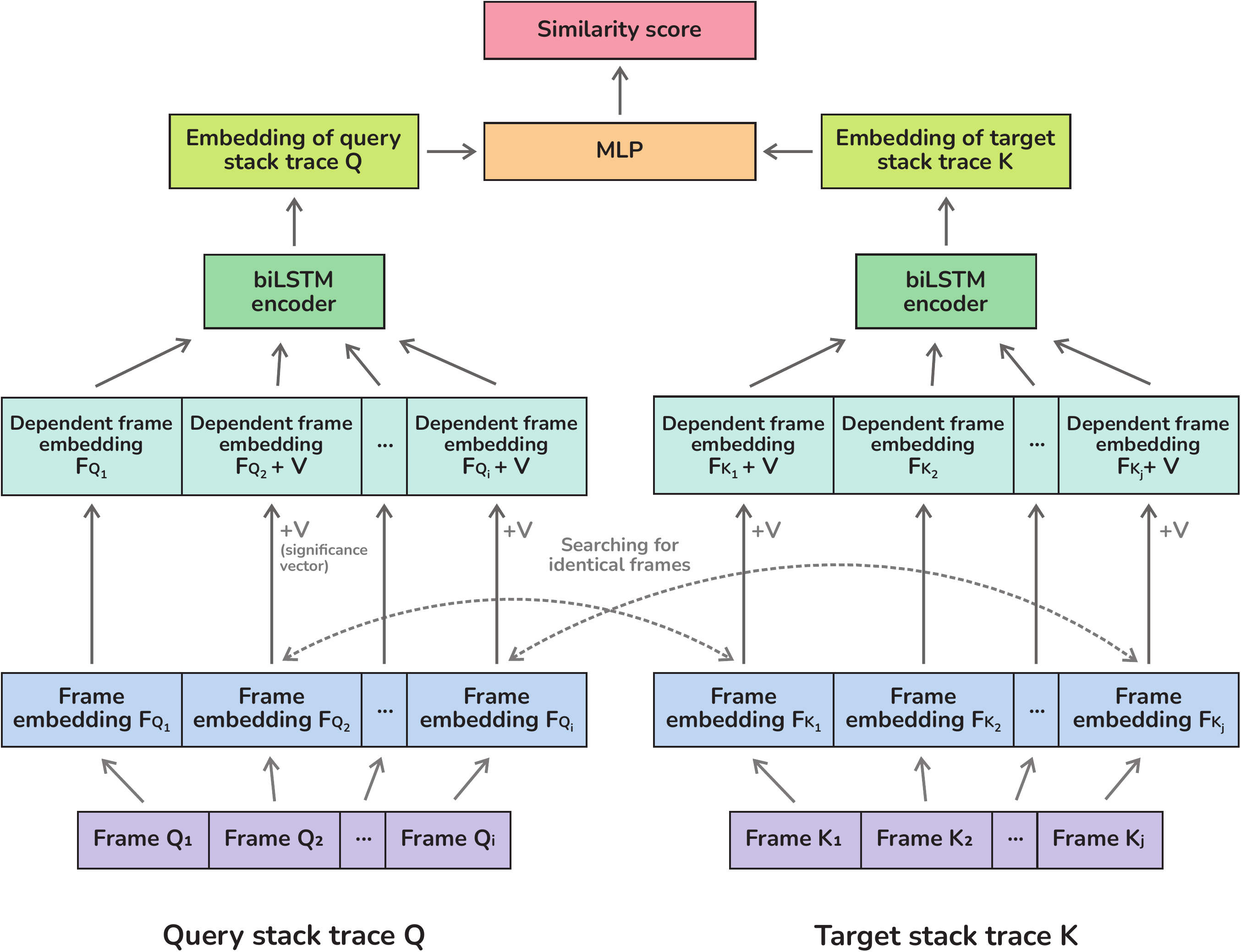}
    \caption{The architecture of the cross-encoder with biLSTM encoder and searching for identical frames. The identical frames are connected by dotted lines.}
    \label{fig:cross_encoder}
    \vspace{-0.2cm}
\end{figure*}

The second stage of our approach is the reranker. It receives \textit{K} most similar stack traces from the embedding model and re-ranks them, taking into account the similar frames between them and the incoming query stack trace. This is an entire separate model, and it serves to enhance the precision of stack trace similarity assessments.

\subsubsection{Preprocessing and tokenization of stack frames}

The preprocessing and tokenization steps for the reranker are identical to those used for the main embedding model, ensuring consistency in handling stack trace data.

\subsubsection{Creating embeddings of stack frames}

The process of converting stack frames to embeddings in the reranker follows the same procedure as described previously, maintaining uniformity in feature extraction across models.

\subsubsection{Creating embeddings of stack traces with cross-encoder}

A cross-encoder is a powerful mechanism in machine learning, specifically designed to process pairs of inputs simultaneously to produce a single output, which is highly beneficial for the retrieval task~\cite{Nogueira2019PassageRW, Humeau2019PolyencodersAA, Qu2019BERTWH, 10.1145/3397271.3401075}. We use a cross-encoder as a reranker in our approach to enhance the precision of our similarity assessments, ensuring that the most relevant stack traces are accurately aligned and compared.

We utilize a cross-encoder model based on a biLSTM architecture to identify identical frames across stack traces. The architecture of the cross-encoder is shown in Figure~\ref{fig:cross_encoder}. This approach begins by encoding each stack frame into a fixed-length vector. If a frame from the query stack trace $Q$, denoted as $Q_i$, is also found in the second stack trace $K$, its representation is enhanced by adding a learned \textit{significance vector}, $V$, which signifies the frame's presence in both traces. 
This procedure converts two independent sequences of frame embeddings, [$F_{Q_1}, F_{Q_2}, ..., F_{Q_i}$] from the first stack trace and [$F_{K_1}, F_{K_2}, ..., F_{K_j}$] from the second, into two interdependent sequences, [$F_{Q_1}, F_{Q_2}+V, ..., F_{Q_i}+V$] and [$F_{K_1}+V, F_{K_2}, ..., F_{K_j}+V$]. Now, the embeddings are enriched with mutual information between corresponding stack frames.

In the next step, a model, equivalent to the embedding model described above, is used to aggregate these two sequences into two embeddings. Then, these embeddings are concatenated and fed into an MLP to obtain a single number --- a similarity score for the input pair.
This method allows the model to emphasize frames that appear in both stack traces, potentially increasing the relevance and accuracy of the similarity score. 
Besides adding a \textit{significance vector} $V$ to the frame embeddings and utilizing it in the final MLP, the two stack traces do not interact directly. Interestingly, even this limited interaction can significantly improve the accuracy of the method.

\subsubsection{Training the cross-encoder}

The training data was constructed using triplets of stack traces: an anchor stack trace, a positive example (a stack trace from the same category), and a negative example (a stack trace from a different category). The positive pairs were sampled as described in Section~\ref{sec:embedding_mode_training}. The negative example was added to each pair by sampling a random stack trace from a different random category.

The Binary Cross-Entropy (BCE) loss was used for training the cross-encoder. The loss function was defined as follows:
\[
L_{\text{BCE}} = \log(1 + \exp(-s_p)) + \log(1 + \exp(s_n))
\]
where \( s_p \) is the similarity score between the anchor and the positive example and \( s_n \) is the similarity score between the anchor and the negative example.
This loss function allows the model to learn to distinguish between similar and dissimilar stack traces effectively, improving the accuracy.

\subsubsection{Final decision}

The reranker processes \textit{K} pairs --- the incoming stack trace \textit{Q} paired with each of the \textit{K} stack traces selected by the embedding model. In then re-ranks these \textit{K} stack traces based on the calculated similarity scores. Finally, the stack trace on top ($R_1$ in Figure~\ref{fig:rerank}) is deemed the most similar one, and its similarity score is compared to threshold~\textit{T}. If it is larger than the threshold, then the incoming stack trace \textit{Q} is placed into the same category as $R_1$, similar to most approaches~\cite{fast, lerch, DURFEX, khvorov2021s3m}. If it is smaller, then no stack trace is similar enough and the new category is created for \textit{Q}.
\section{Evaluation}
\label{sec:evaluation}

\begin{table*}[]
\centering

\caption{Comparison of datasets.}

\begin{tabular}{lrrrrr}
\toprule
\multirow{2}{*}{\textbf{Dataset}}& \textbf{Number of} & \textbf{Number of}  & \textbf{Number of} &\textbf{Average number of} &\textbf{Average number of}  \\
                              & \textbf{reports} &  \textbf{unique reports} & \textbf{categories} & \textbf{reports in category} & \textbf{unique reports in category}  \\ \midrule
Ubuntu & 15,293 & 9,792 & 3,825 & 3.99 & 2.77  \\ 
Eclipse & 55,968 & 47,901 & 47,636 & 1.17 & 1.13  \\ 
NetBeans & 65,417 & 49,842 & 51,714 & 1.26 & 1.17  \\ 
Gnome & 218,160 & 57,553 & 100,944 & 2.16  & 1.29 \\ \midrule
SlowOps & 886,730 & 66,451 & 1,361 & 651.52 & 49.81  \\ \bottomrule
\end{tabular}

\vspace{-0.3cm}
\label{tab:datasets}
\end{table*}

This section presents the multi-faceted evaluation that we conducted of our approach, as well as the existing state-of-the-art approaches. Striving for evaluating the performance needed in practice, we posed the following research questions:

\begin{enumerate}[font={\bfseries},label={RQ\arabic*},leftmargin=1.2cm]
    \item How accurate is our approach in predicting category for the given error report?
    \item How well can our model distinguish between a situation when error reports are attached to the existing category and when a new category is created? 
    \item How fast is our method when incorporated into the retrieval process?

\end{enumerate}

Below, we describe the datasets we used, including the new one we present, the evaluation setup, the baselines, the metrics, and the results of our experiments.

\subsection{Datasets}
\label{sec:datasets}
\subsubsection{Existing datasets}
\label{subsec:exdatasets}

The majority of evaluations presented in previous works utilize the following public datasets: Ubuntu, Eclipse, NetBeans, and Gnome, which were released by Campbell et al.~\cite{campbell2016unreasonable} and Rodrigues et al.~\cite{rodrigues2022tracesim}. These datasets compile reports from bug tracking systems across the respective projects. Most applications within Gnome and Ubuntu are developed in C/C++, while Eclipse and NetBeans are two popular Integrated Development Environments (IDEs) implemented using Java. 

The datasets contain crash reports with detailed stack traces, timestamps, and manually assigned categories. The data indicates whether each report is a duplicate or corresponds to a new unique category. The manual labeling by developers acts as ground truth because it reflects human judgment and expertise, ensuring accurate categorization for evaluating deduplication methods.

A significant issue with these datasets is the limited size of the categories. As can be seen in Table~\ref{tab:datasets}, the datasets contain many categories, each with only a single report. This limitation can hinder the evaluation of a similarity model, as it provides minimal information about each category, thus affecting the robustness of the assessment. In addition, it is interesting to see whether proprietary data follows the same pattern.

Note also that the \textit{average number of unique reports in category} in Table~\ref{tab:datasets} is not equal to simply the \textit{number of unique reports} divided by the \textit{number of categories}. This is because there are situations where two different categories contain identical reports. This may stem from different reasons, including creating categories based on something other than the stack trace itself, \textit{e.g.}, OS or product versions. Most of the existing approaches, as well as ours, do not take this metadata into account, however, we take the presence of the same reports in different categories into consideration when designing our experiments, as described further in Section~\ref{sec:evaliuation_setup}.

\subsubsection{\dataset}

To address these issues, we present \dataset, a new industrial dataset that contains stack traces from JetBrains, a large vendor of tools for software developers and teams. The dataset contains stack traces from IntelliJ-based products~\cite{kurbatova2021intellij} that arrived from 26.01.2021 to 29.02.2024. 
More specifically, the dataset comprises reports triggered by \textit{Slow Operation Assertion}. This error occurs when an action in the UI thread takes longer than a specified amount of time to complete. Consequently, these reports assist in identifying not logical errors within the application, but performance issues encountered by the user. Nevertheless, the approach for addressing these reports is identical to that used for typical logical errors, with the main distinction being that the most informative frames for localizing the error may not be at the top of the trace but rather in the middle. The distinctive nature of this dataset makes it an interesting candidate for comparison with more general open-source ones.

\dataset was preprocessed through the following steps:

\begin{enumerate}
    \item Categories were selected only if the issue was manually reviewed by a JetBrains developer. \textit{Slow Operation Assertion} errors differ by a particular combination of frames where the UI thread calls some long operation, and so developers manually created categories for unique combinations of these frames.
    \item Reports containing calls to third-party plugins, unrelated to the platform's codebase, were removed to ensure the privacy of the users, whose identity might be revealed through the names of plugins. This also helps to keep the dataset self-contained. Approximately 30\% of reports were removed after this step.
\end{enumerate}

In Table \ref{tab:datasets}, one can see that \dataset differs significantly from the existing datasets in several key aspects. Unlike the previous datasets, which contain a high number of categories with only a single error report, our dataset includes a large number of reports per category, with an average of 49.8 unique reports per category. UI-related issues happen a lot, and with tens of thousands of users, this allows us to collect a large number of different crashes with the same underlying issue. This ensures that the dataset is more robust and provides a richer source of information for assessing the performance of similarity models. It is also much less prone to the issue of having identical reports in different categories, ensured by the company's data collection systems.

By introducing this new dataset, we offer a different perspective that can be valuable for evaluating models on this task. While it might not generalize, this diversity in datasets is crucial as it allows for a more comprehensive assessment of model performance, ensuring that models are not only effective in traditional scenarios but also adaptable to contemporary real-world applications.

\subsection{Evaluation Procedure}
\label{sec:evaliuation_setup}

Our evaluation procedure aims to model real-world scenarios. Each dataset is first sorted by the arrival time of error reports and then split into training, validation, and test sets in the proportions of 70/10/20 based on the number of reports.

All reports from the test split of the dataset are sorted by their arrival time. We then iterate over all reports in the test segment and use the similarity model to rank categories for each new report. Specifically, we compute similarities between each previously arrived report and the newly arrived one. For each category, a similarity score is calculated based on the highest similarity score among its stack traces:
$$
\text{Similarity} (q, \text{category}) = \max_{k \in \text{category}} \text{Similarity}(q, k)
$$

\noindent where \(q\) represents the current stack trace, and \(k\) is a stack trace within the category. This part is common in the evaluation, because in all the compared approaches, the incoming report is assigned to the category that has the single most similar report to the incoming one. Based on this ranking of categories, we compute the metrics presented in Section \ref{sec:metric}.

A significant change we made compared to the previous evaluation procedure used in the work by Khvorov et al. \cite{khvorov2021s3m} is that if, during the iteration over reports in the test split, we encounter a report that already has an \textit{identical} report in an existing category, we ignore this report in the evaluation process. There are two main reasons for this adjustment.

 Firstly, in real-world scenarios, \textit{e.g.}, at JetBrains, when a report arrives identical to an existing one, it is automatically added to the same category without even employing the similarity model. Thus, it does not make sense to evaluate the model in situations where it would not be used.
 
Secondly, as described in Section~\ref{sec:datasets}, in some datasets, there are situations where two different categories contain identical reports. This can unfairly influence the attach metric, as the top-ranked category would be one of these categories, regardless of the performance of the similarity model.

In all our experiments, we use the following hyperparameters for our approach:

\begin{itemize}
    \item \textbf{Size of BPE vocabulary} is 10,000 tokens. We leave detailed experiments with this parameter for future work.
    \item The \textbf{number of stack traces that the Embedding model selects} and passes to the Reranker $K = 10$. We conducted preliminary experiments with different values, from 5 to 100, and found that the larger ones provide negligible improvements while costing time. Still, we leave mode details experiments for future work.
    \item The \textbf{threshold} that decides if the best stack trace is similar enough to move the incoming one to it or create a new category $T$ is unique for each dataset and is defined as the one which results in the best F1 score. The logic here is as follows. This problem can be viewed as a binary classification task where the objective is to determine whether a given report should create a new category. Setting the threshold to 0 would result in never creating new categories, while setting the threshold to 1 would result in creating a new category for each new stack trace that is not exactly identical to an existing one. Having the historical data about the creations of categories, we can evaluate different thresholds to find the balance using F1 score. This way, the threshold is trained for each particular dataset, since Table~\ref{tab:datasets} shows just how different their structure is.
\end{itemize}

\subsection{Baselines}

We consider several baselines for comparison with our proposed method, categorized into supervised, unsupervised, and large language models (LLMs). Below, we detail the specific models used in each category.

\subsubsection{Supervised baselines}

Supervised methods require training data to learn and typically involve models that adjust their parameters based on annotated datasets to enhance their predictive accuracy. Our solution falls into this category.

\textbf{S3M} \cite{khvorov2021s3m}: S3M is the first deep learning model proposed by Khvorov et al. At the time of its introduction, it outperformed non-deep learning solutions and was considered a state-of-the-art approach. It utilizes a biLSTM model to encode stack traces, but does not employ BPE, FAISS, or reranking.

\textbf{DeepCrash} \cite{deepcrach}: Proposed by Liu et al., DeepCrash is another supervised solution with an advanced architecture that outperforms S3M. The core ideas of the paper are to switch to the embedding model and use skip-gram for obtaining embeddings of the stack frames.

\subsubsection{Unsupervised baselines}

Unsupervised models do not require labeled data and are advantageous in scenarios where such data is unavailable. However, they still rely on unlabeled data, such as a large collection of reports, to train components like IDF (Inverse Document Frequency).

\textbf{Lerch} \cite{lerch}: The model proposed by Lerch and Mezini, employing TF-IDF scores as its base.

\textbf{FaST} \cite{fast}: FaST is an unsupervised solution presented by Rodrigues et al. The approach focuses on working in linear time by considering frames at the top of stack traces.

\subsubsection{Large language models (LLMs)}

LLMs are a category of unsupervised approaches trained on vast corpora of text data, incorporating extensive information. 

\textbf{text-embedding-3-small}: This LLM for creating embeddings by OpenAI \cite{openai2022textembedding} is chosen for comparison due to its strong performance. We selected the smaller version as it outperformed the larger ``text-embedding-3-large'' in our tests. When working with an LLM, we concatenate all frames in a stack trace into a single string using a delimiter, and pass this string to the LLM. After receiving an embedding, we work with it in the same manner as in our approach.

Each of these models provides a different perspective on the task of error report categorization, and by comparing our approach with these baselines, we aim to demonstrate its effectiveness and versatility.

    \begin{table*}[]
\caption{RQ1: Acc@1 for all models on all datasets, measuring the accuracy of assigning incoming reports into the correct category.}
\centering
\begin{tabular}{lcccc|c}
\toprule
                            & \multicolumn{1}{l}{\textbf{Ubuntu}} & \multicolumn{1}{l}{\textbf{Eclipse}} & \multicolumn{1}{l}{\textbf{NetBeans}}  & \multicolumn{1}{l|}{\textbf{Gnome}} & \textbf{SlowOps} \\ \midrule
S3M & 0.32 & 0.52 & 0.25  & 0.25 & 0.95\\ 
DeepCrash & 0.39 & 0.10 & 0.32  & 0.36 & 0.96 \\ 
Lerch & 0.40 & 0.60 & 0.22  & 0.37 & 0.96\\
FaST & 0.41 & 0.72 & 0.30  & 0.32 & 0.97 \\ 
text-embedding-3-small & 0.62 & 0.73 & 0.46  & 0.40 & 0.93 \\ \midrule
Ours (Embedding model only) & 0.61 & 0.71 & 0.50  & 0.44 & \multicolumn{1}{c}{\textbf{0.98}} \\
Ours (Embedding model + Reranker) & \textbf{0.65} & \textbf{0.75} & \textbf{0.52}  & \textbf{0.45} & \multicolumn{1}{c}{\textbf{0.98}} \\ \bottomrule
\end{tabular}

\label{tab:acc1}
\end{table*}

\begin{table*}[]
\centering
\caption{RQ2: ROC-AUC for all models on all datasets, measuring the ability to correctly create new categories when needed.}
\begin{tabular}{
lcccc|c }
\toprule
 &  \textbf{Ubuntu} &  \textbf{Eclipse} &  \textbf{NetBeans}  &  \textbf{Gnome} &  \textbf{SlowOps} \\ \midrule
 S3M &  0.51 & 0.73 &  0.55  & 0.60 &  0.81 \\ 
 DeepCrash &  0.64 & 0.64 &  0.62  &  0.62 &  0.98 \\ 
 Lerch &  0.56 &  0.79 &  0.55  &  0.61 & 0.85 \\
 FaST & 0.62 &  0.67 &  0.58  &  0.60 & 0.95 \\ 
 text-embedding-3-small & 0.75 &  0.85 & 0.67  &  0.68 &  \textbf{0.99} \\\midrule
 Ours (Embedding model only) &  \textbf{0.76} &  0.85 & 0.71  & 0.70 &  \textbf{0.99} \\ 
 Ours (Embedding model + Reranker) &  0.75 &  \textbf{0.86} &  \textbf{0.72}  & \textbf{0.71} &  0.96 \\ 

 \bottomrule
\end{tabular}

\label{tab:roc_auc}
\end{table*}

\begin{table}[]
\centering
\caption{RQ3: Speed comparison of all models on the Ubuntu dataset. The time is the average per one report.}
\begin{tabular}{lr}
\toprule
                            & \textbf{Time,  ms}       \\ \midrule
S3M                         & 1722.8          \\
DeepCrash                   & \textbf{7.6}    \\
Lerch                       & 307.4           \\
FaST                        & 916.2           \\ 
text-embedding-3-small      & 1021.2          \\\midrule
Ours (Embedding model only) & 8.7             \\ 
Ours (Embedding model + Reranker)        & 144.5           \\ 
 \bottomrule
\end{tabular}

\label{tab:speed}

\end{table}

\subsection{Metrics}
\label{sec:metric}

\subsubsection{RQ1: Attach accuracy metric}

To address RQ1 and study the general accuracy of our approach in choosing the necessary category for incoming stack traces, we employ the Acc@1 metric. In our evaluation, there are two types of reports: those that are attached to an existing category and those that create a new category. The Acc@1 metric is applicable only to reports that are attached to some existing category, while the correctness of creating new categories is studied in RQ2.
Acc@1 measures the ratio of cases where the model correctly predicts the most suitable category for a given report:
\[
\text{Acc@1} = \dfrac{\sum_{r \in A}[\text{Most relevant category predicted}]}{\|A\|}
\]
where \(A\) is the set of attached reports, and \(r\) is one such report.

\subsubsection{RQ2: Correctness of creating new categories}

To address RQ2 and study how well the models can decide that the new category must be introduced, we use the ROC-AUC score, following Rodriguez et al.~\cite{fast}.
As we mentioned in Section~\ref{sec:evaliuation_setup}, we treat this problem as a binary classification task and our approach uses the threshold \textit{T} that balances the precision and recall of this binary classification using an F1-score. Evaluating this task using ROC-AUC allows to compare the models in general, before selecting the threshold.

\subsubsection{RQ3: Speed metric}

To address RQ3, we measure the time required for the model to compute the similarity scores for all the necessary reports. Specifically, the time taken to calculate the embeddings for deep learning based approaches is not taken into account, because in large software systems they are pre-computed from the previous runs of the model. Since the incoming report gets sent to the category with the single most similar report (or creates a new category), the whole process involves only finding this most similar report, which is precisely what we are measuring. For our model, we distinguish between the time taken with the reranker and without it. Since this measurement just scales with the size of the dataset and the models differ by orders of magnitude, we measure this RQ on just one dataset, Ubuntu.

\subsection{Results}

\subsubsection{RQ1: Assigning categories}

As shown in Table \ref{tab:acc1}, our method outperforms all baselines across all datasets. Specifically, on the Eclipse and Ubuntu datasets, our method with the reranker achieves Acc@1 scores of 0.75 and 0.65, respectively, which are higher than the closest competitor, the text-embedding-3-small model. It can also be seen that on open-source datasets, text-embedding-3-small is the strongest among other models, showcasing the power of LLMs. Finally, it is evident that the results on \dataset are much more positive than on open-source ones. This can be due to the specific nature of the issues in it or due to many more reports per category, which ensures a more robust classification. In any case, this highlights how important it is to evaluate the models on different kinds of data.

\subsubsection{RQ2: Creating new categories}

In terms of ROC-AUC, which measures the ability to distinguish when a new category should be created, our method again outperforms most baselines, as shown in Table \ref{tab:roc_auc}. On the Eclipse dataset, our approach with the reranker achieves a ROC-AUC of 0.86, and on the Gnome dataset, it reaches 0.71. Notably, our method without the reranker achieves the highest ROC-AUC score of 0.99 on our dataset, suggesting that even without reranking, our model is highly effective at identifying new categories. Again, the second-best model is text-embedding-3-small, demonstrating excellent results.

\subsubsection{RQ3: Speed}

Finally, as detailed in Table \ref{tab:speed}, the speed of our model indicates a trade-off between accuracy and computation time. While our method with the reranker provides the best predictive performance, it requires 144.5 ms per report, which is an order of magnitude faster than some baselines like S3M (1722.8 ms) but similarly an order of magnitude slower than others like DeepCrash (7.6 ms). However, when using only the embedding model without reranking, our method is significantly faster, processing each report in just 8.7 ms, making it suitable for real-time applications where speed is critical. While DeepCrash is excellent in terms of time, Tables~\ref{tab:acc1} and~\ref{tab:roc_auc} show that it struggles with performance. In contrast, text-embedding-3-small, which shows great results, is very slow, which is to be expected of an LLM. 
\section{Discussion}
\label{sec:discussion}

\subsection{Balance Between Accuracy and Performance}

Overall, our approach can be seen as a great practical compromise that can perform in both accuracy and time. The two-staged architecture --- comprising an embedding model followed by a reranker --- proves to be effective, with the reranker enhancing the precision of category predictions. This approach improves upon the challenges identified in previous works, and it can still perform well without the reranker if time is an issue in a particular task.

Other evaluated models demonstrate very different performance. The tested LLM showed very good accuracy, close to our approach, but it is really slow. Two fastest approaches disregarding ours, although on a very different scale, --- are DeepCrash and the approach by Lerch and Mezini, however, their accuracy is significantly worse. It is thus important to continue research in this area, but taking into account the practical aspects and ensuring that new ``state-of-the-art'' approaches are good not only in base accuracy.

\subsection{Large Language Models (LLMs)}

Given their entirely different nature and importance, it is worth discussing the LLMs separately. While LLMs like the text-embedding-3-small model perform well in our evaluations, they have certain limitations. They often require making an API call, which can introduce latency, making them less suitable for real-time applications. Additionally, software companies may worry about their privacy and may not be willing to send the data over the internet to the most powerful models. While local-based LLMs can be used to overcome these issues, running a biLSTM locally is usually cheaper and easier, thus, our model offers a more tailored solution for this specific task. Nonetheless, the provided results clearly indicate that LLMs deserve further study in this field.

\subsection{Performance on \dataset}

An interesting result from our evaluation is just how different the results are on our new industrial dataset. The reports in this dataset all relate to one error type, \textit{Slow Operation Assertion}, and the dataset has an order of magnitude more reports per category, which might explain why it is ``easier'' for the models to correctly classify them. Still, the good performance of the models on it is not a reason for complacency, because it represents just one specific data source. Rather, this difference of results indicates the importance of carrying out the evaluation on \textit{diverse} data, to obtain a more exhaustive picture of how models will perform in different scenarios.

\section{Threats to Validity}
\label{sec:ttv}

The large-scale nature of our work makes it subject to the following threats to validity.

\textbf{Hyperparameters}. When evaluating our approach, we used certain hyperparameters, such as the size of BPE vocabulary of 10,000 and the number of the most similar stack traces passed from the Embedding model to the Reranker $K = 10$. While we carried out some preliminary experiments to select them, we did not conduct exhaustive evaluations, and so these hyperparameters might not be optimal. Such evaluations are a part of our future work.

\textbf{Model generalization}. Our model is tailored to the dataset it was trained on, which could lead to overfitting to the specific characteristics of that dataset. To address this, we tested the model on a variety of datasets from different sources, helping to ensure its robustness. However, generalizability may still vary in significantly different environments.

\textbf{Data source bias}. The effectiveness of our model is closely tied to the specific dataset used for its training and evaluation. Since \dataset was obtained from a particular software company, JetBrains, it may not fully represent the variability found in other environments, such as open-source projects or smaller companies. The characteristics of stack traces, coding practices, and error reporting can vary significantly across different domains, potentially leading to different outcomes if the model is applied outside the context of our dataset. For this reason, we encourage further researchers to collect even more diverse datasets and share them with the community.

\textbf{Language and framework dependency}.
The stack traces used in our study come primarily from JVM-based languages and C++. This selection might influence the performance of our model when applied to other programming languages or frameworks. The structure of stack traces and the nature of errors can vary significantly depending on the programming environment, potentially affecting the accuracy and effectiveness of our approach. Future work is necessary to explore the model's adaptability to a broader range of programming languages and development environments.

While these threats to validity are important to acknowledge, we believe they do not invalidate the overall conclusions of our study or its practical relevance.
\section{Conclusion}
\label{sec:conclusion}

In this work, we presented a novel model for deduplicating stack traces. The proposed approach consists of two steps: a base embedding model that employs the approximate nearest neighbors search to quickly find the most similar stack traces and a reranker that re-ranks them more accurately, taking into account the information about repeated individual frames. To facilitate more detailed comparison of our model with existing approaches, we collected a dataset of stack traces from JetBrains, complimenting the existing open-source datasets. Finally, we carried out a multi-faceted evaluation, comparing the accuracy of categorization, but also evaluating the ability of models to create new categories and their operation time. Our approach shows the best results in terms of accuracy and is efficient in terms of time. We release the dataset~\cite{dataset} and the code~\cite{replication} to the community to facilitate further research.

In the future work, we plan to continue to improve our approach by conducting more experiments with different hyperparameters and architectures. We also want to explore even broader evaluations, considering different types of industrial projects, languages, and frameworks. We believe this is crucial for ensuring the applicability of the research in practice.

\bibliographystyle{IEEEtran}
\balance
\bibliography{IEEEabrv,paper}

\end{document}